# Nonreciprocal electrical transport in multiferroic semiconductor (Ge,Mn)Te

Ryutaro Yoshimi[1*], Minoru Kawamura[1], Kenji Yasuda[1,2], Atsushi Tsukazaki[3], Kei S. Takahashi[1],

Masashi Kawasaki[1,4], and Yoshinori Tokura[1,4,5]

[1] RIKEN Center for Emergent Matter Science (CEMS), Wako 351-0198, Japan.

[2] Department of Physics, Massachusetts Institute of Technology, Cambridge, Massachusetts

02139, USA

[3] Institute for Materials Research, Tohoku University, Sendai 980-8577, Japan

[4] Department of Applied Physics and Quantum-Phase Electronics Center (QPEC),

University of Tokyo, Tokyo 113-8656, Japan

[5] Tokyo College, University of Tokyo, Tokyo 113-8656, Japan

* Corresponding author: ryutaro.yoshimi@riken.jp

# ABSTRACT

We have investigated the nonreciprocal electrical transport, that is a nonlinear resistance effect depending on the current direction, in multiferroic Rashba semiconductor (Ge,Mn)Te. Due to coexistence of ferromagnetic and ferroelectric orders, (Ge,Mn)Te provides a unique platform for exploring the nonreciprocal electrical transport in a bulk form. (Ge,Mn)Te thin films shows a large nonreciprocal resistance compared to GeTe, the nonmagnetic counterpart with the same crystal structure. The magnetic-field-angle dependence of the nonreciprocal resistance is maximized when magnetic field is orthogonal to both current and electric polarization, in accord with the symmetry argument. From the analysis of temperature and magnetic field dependence, we deduce that inelastic scatterings of electrons mediated by magnons dominantly contribute to the observed nonreciprocal response. Furthermore, the nonreciprocal resistance is significantly enhanced by lowering hole density. The Fermi level dependence is attributed to the deformation of the Rashba band in which the spin-momentum locked single Fermi surface appears by exchange field from the in-plane magnetization. The present study provides a key insight to the mechanisms of novel transport phenomena caused by the interplay of ferroelectric and ferromagnetic orders in a semiconductor.

**MAIN TEXT**

In materials systems without spatial inversion symmetry, electronic energy bands exhibit momentum-dependent spin-splitting by spin-orbit coupling. The Rashba spin-split bands in polar materials are one such example [1,2], which appears in various materials systems such as heterointerfaces of semiconductors, surfaces of heavy metals, and bulk polar semiconductors [3–7]. They have provided a materials platform to study novel transport and spintronic properties including the spin Hall effect and the spin galvanic effect [8–10].

Nonreciprocal electrical transport, in which the electrical resistance depends on the current direction, is one of the characteristic transport phenomena in inversion broken electronic systems [11,12]. In polar electronic structure with electric polarization $\boldsymbol{P}$, the electronic resistance that is characterized by magnetic field $\boldsymbol{B}$ and current $\boldsymbol{I}$ is expressed by,

$$R(\boldsymbol{B},\boldsymbol{I}) = R_0[1 + \beta \boldsymbol{B}^2 + \gamma(\boldsymbol{I} \times \boldsymbol{B}) \cdot \widehat{\boldsymbol{P}}] \tag{1}$$

where the first and second terms represent the resistance at zero magnetic fields and normal magnetoresistance characterized by $R_0$ and $\beta$, respectively. $\widehat{\boldsymbol{P}}$ represents the unit vector pointing to $\boldsymbol{P}$. The third term which depends on both $\boldsymbol{I}$ and $\boldsymbol{B}$ is the nonreciprocal resistance, also termed as unidirectional magnetoresistance. As is clear from the expression, the nonreciprocal response is a nonlinear effect with respect to current, and allowed by symmetry when $\boldsymbol{P}$, $\boldsymbol{B}$ (or $\boldsymbol{M}$) and $\boldsymbol{I}$ are orthogonal to each other. The coefficient $\gamma$ characterizes the nonreciprocal electrical transport, which is the ratio of nonreciprocal resistance and normal resistance $R_0$. Recently, large nonreciprocal responses have been observed in surface/interface systems with ferromagnetism, such as heavy metal/ferromagnetic-metal bilayer interface and magnetic topological insulator surface [13–16]. A Rashba semiconductor doped with magnetic elements would also be a unique platform to explore the nonreciprocal electrical transport. The introduction of the ferromagnetism is expected to largely deform the Rashba band structure, giving a knob for controlling the nonreciprocal transport. However, the nonreciprocal transport in bulk magnetic Rashba systems has not been studied so far, possibly due to a limited number of conducting materials that show ferroelectricity and ferromagnetism simultaneously.

Mn-doped GeTe (crystal structure shown in Fig. 1(a)) is a multiferroic Rashba semiconductor in which both ferromagnetic and ferroelectric orders appear in a bulk form. The parent compound GeTe is a polar semiconductor with p-type carriers in which inversion symmetry is broken by the displacement of Ge and Te ions along [111] direction below $T_\text{C}^\text{FE} \sim 700$ K [7,17–21]. The polar crystal symmetry of GeTe results in the large Rashba-type spin splitting of the bulk band structure, whose Rashba parameter $\alpha_\text{R}$ is as large as 2-4 eVÅ [7,22]. By partially substituting Ge with Mn, a ferromagnetic order shows up while keeping the polar crystal symmetry [22–24]. The ferromagnetic transition temperature depends on Mn content and hole density, which can be as high as 180 K [23,24]. The coexistence of ferromagnetism and the polar symmetry is demonstrated by angular-resolved photoemission spectroscopy [22].

Here, we investigate the nonreciprocal electrical transport in multiferroic Rashba semiconductor (Ge,Mn)Te thin films. The nonreciprocal resistance in (Ge,Mn)Te is confirmed to be much larger than that in GeTe, suggesting a crucial role of the exchange field effect powered by ferromagnetism. The nonreciprocal resistance is maximized when magnetization is perpendicular to both magnetic-field and electric-current directions, and its sign is switched upon the reversal of magnetization, in accord with the symmetry argument of this multiferroic conductor. The observed nonreciprocal resistance under variations of applied magnetic field and chemical potentials suggests a crucial role of asymmetry in magnon-mediated scatterings in the uniquely spin-momentum-locked electron system.

We grew 75-nm thick (Ge,Mn)Te films on InP(111)A substrates by molecular beam epitaxy (MBE). The epi-ready substrates were annealed at 380 ºC at a base vacuum pressure of approximately $1 \times 10^{-7}$ Pa before the deposition of thin films at 200 °C. We inserted $Sb_2Te_3$ (1 nm) and GeTe (1 nm) buffer layers beneath the (Ge,Mn)Te layer to stabilize the rhombohedral distortion of (Ge,Mn)Te [25]. The orientation of the (Ge,Mn)Te film is along [111] direction as confirmed by x-ray diffraction [25], which is parallel to the ferroelectric polarization. It is reported that the epitaxial GeTe film grown via MBE technique is single domain over several hundred µm² [7,26]. The equivalent pressures of beam flux for Ge and Mn were $P_{Ge} = 4.5 \times 10^{-6}$ Pa and $P_{Mn} = 5.0 \times 10^{-7}$ Pa, respectively. We evaluated the actual composition of Mn in (Ge,Mn)Te by inductively coupled plasma mass spectroscopy and found it to be 9 %. The growth duration for (Ge,Mn)Te layer was 30 minutes, corresponding to the growth rate of approximately 2.5 nm per minute. After the thin film growth, we capped the sample with $AlO_x$ by the atomic layer deposition method to prevent the sample from degradation, and then defined Hall bar devices with 10 µm in width by UV photolithography and Ar ion milling. The current direction of the Hall bar devise is along $\bar{\Gamma} - \bar{K}$ direction on the surface brillouin zone [6]. The length between voltage probes in the Hall bar devices was 10 µm. The electrodes were Ti (5 nm) and Au (25 nm) formed by electron beam deposition.

We evaluated the nonreciprocal electrical transport properties as a response to ac electric current. We applied ac electric current $I(t) = \sqrt{2}I_0 \sin(\omega t)$ and measured the quadrature component of the second harmonic voltage $V_{xx}^{2\omega}$ by lock-in technique. Since the electrical polarization of (Ge,Mn)Te is parallel to the out-of-plane direction ($P \parallel z$), the nonlinear resistance $R_{xx}^{2\omega}$ is to be maximized under the $I \parallel x$ and $B \parallel y$ configuration (Fig. 1(b)). We anti-symmetrize $V_{xx}^{2\omega}$ with respect to $B$ and define nonreciprocal resistance as $R_{xx}^{2\omega} = V_{xx}^{2\omega}/I_0 = -\frac{1}{\sqrt{2}}R_0 \gamma B I_0$ (see Eq. (1)). We performed the electrical transport measurement using Physical Properties Measurement System (PPMS, Quantum Design) equipped with a rotation probe to tilt the sample with respect to the magnetic field. For the nonlinear measurement, we used a current source Model 6221 (Keithley), a lock-in amplifier LI5650 (NF Corporation), and signal amplifier SR560 (Stanford Research Systems). The measurement frequency $f = \omega/2\pi$ was set to 11 Hz.

Figure 1(c) shows the in-plane magnetic field $B_y$ dependence of the nonreciprocal resistance $R_{xx}^{2\omega}$ at $T = 2$ K for (Ge,Mn)Te with hole density $p = 1.8 \times 10^{19}$ cm$^{-3}$ and for a pristine 70-nm thick GeTe film with $p = 9.4 \times 10^{19}$ cm$^{-3}$. The excitation current is 10 µA for (Ge,Mn)Te and 100 µA for GeTe. The magnetic anisotropy of (Ge,Mn)Te with this Mn content is small; the magnetization can be fully aligned to the direction of $B_y$, that is in-plane direction, at 0.5 T [23]. $R_{xx}^{2\omega}$ in the pristine GeTe is negligibly small over a wide range of magnetic field; the evaluation of nonreciprocal coefficient in GeTe by applying larger current is shown in Supplementary Materials (SM). Meanwhile, a sizable $R_{xx}^{2\omega}$ signal appears in (Ge,Mn)Te showing a unique field dependence. With the increase of $B_y$, $R_{xx}^{2\omega}$ at first steeply increases below $B_y \sim 0.8$ T, and then rather decreases with further increasing $B_y$. The significant difference between (Ge,Mn)Te and GeTe indicates an important role of ferromagnetism in nonreciprocal electrical transport, *i.e.*, Zeeman-like effect arising from the exchange field produced by ferromagnetically ordered Mn moments. Figure 1(d) shows the current amplitude dependence of $R_{xx}^{2\omega}$ at $B_y = 0.8$ T. $R_{xx}^{2\omega}$ linearly increases in a low current region and deviates from the linear dependence in the higher current region; magnetic field dependence of $R_{xx}^{2\omega}$ with different excitation current is shown in Fig. S2 in Supplementary Materials (SM). We define the coefficient of nonreciprocal electrical transport $\gamma$ (see Eq. (1)) in the low current region and $B_y < 0.8$ T. Here, we use $\gamma' = \gamma S$, where $S$ represents the cross-sectional area of the sample that is 75 nm × 10 µm, to eliminate the sample size dependence [27]. In the present sample, $\gamma' = 1.3 \times 10^{-11}$ m$^2$T$^{-1}$A$^{-1}$, which is five orders of magnitude larger than pristine GeTe (see Section S1 in SM and Ref. [28]).

Figure 2(a) shows the magnetic field dependence of $R_{xx}^{2\omega}$ for several azimuth angles $\varphi$ (defined as the inset of Fig. 2(b)). The absolute value of $R_{xx}^{2\omega}$ is the largest at $\varphi = 0°$ ($B \parallel y$) and $180°$ ($B \parallel -y$), while the sign is the opposite. In addition, $R_{xx}^{2\omega}$ is negligibly small at $\varphi = 90°$ ($B \parallel x$). We plot the $\varphi$ dependence of $R_{xx}^{2\omega}$ at $|B| = 0.8$ T (Fig. 2(b)), which is well fit with the cosine curve (shown in a black line). A similar angular dependence appears with respect to $\theta$, that is a tilt angle of the magnetic field from the $z$-axis to the $y$-axis (Fig. 2(c)). From these magnetic-field direction dependences, it turned out that $R_{xx}^{2\omega}$ is the largest when $B$ is perpendicular to both electric current and electric polarization, as expected from Eq. (1).

In the following, we consider the band structure of (Ge,Mn)Te to identify a microscopic origin for the observed nonreciprocal electrical transport. We employ the following model Hamiltonian that represents a Rashba semiconductor with magnetization $M$,

$$H = \frac{\hbar^2 k^2}{2m^*} + \alpha(k_x \sigma_y - k_y \sigma_x) + \Delta \hat{M} \cdot \boldsymbol{\sigma} \tag{2}$$

where $\hbar, \boldsymbol{\sigma} = (\sigma_x, \sigma_y, \sigma_z), m^*, \alpha, \Delta$, and $\hat{M}$ respectively represent Planck's constant, Pauli matrices, the effective mass, the Rashba parameter, the parameter for the exchange interaction, and the unit vector pointing to magnetization. Using the band parameters $m^* = 0.6\ m_0$, $\alpha = 2.0$ eVÅ, and $\Delta = 50$ meV, where $m_0$ represents the mass of a free electron, we describe the band structure of (Ge,Mn)Te near the band edge obtained by the photoemission spectroscopy [22] (Fig. 3(a)). The characteristic structures of the magnetic Rashba band, such as the spin-splitting of the quadratic band and the exchange gap at the degeneracy point, are reproduced. A band structure with in-plane magnetization ($M \parallel y$) is calculated with the same band parameters, where the two branches of Rashba bands energetically shift in the opposite direction by $2\Delta$ (Fig. 3(b)). Using the model Hamiltonian, we deduce that the hole density $p = 1.8 \times 10^{19}$ cm$^{-3}$ of the present sample corresponds to the position of $E_F$ close to the band edge (see a broken line in Fig. 3(b) and section S4 in SM), where an isolated spin-momentum-locked Fermi surface exists.

In such a single spin-momentum-locked Fermi surface, the spin scattering of conduction electron associated with magnon can contribute to resistance. Without magnon excitations, the difference in the spin angular momentum between the initial and final states would prohibit the backward scattering of electrons. However, by transferring the difference of angular momentum to magnons (Fig. 3(c)), the back scatterings are allowed and thus contribute to electrical resistance. The asymmetry in the probability between magnon absorption and emission processes can produce the nonreciprocity of resistance. In fact, the magnon-mediated nonreciprocal resistance was previously discussed in the case of a magnetic topological insulator bilayer which also has a single spin-momentum-locked Fermi surface in the surface Dirac band [15]. In the present experiment, the nonreciprocal resistance (Fig. 1(c)) decreases in the high field regime (similar magnetic field dependences are observed in other samples as shown in Fig. S5 in SM). This behavior can be explained by the decrease in the population of the magnons as the magnon excitation gap increases with the magnetic field.

To further verify the magnon-mediated scattering mechanism, we examined the temperature dependence of $R_{xx}^{2\omega}$ under several magnetic fields (Fig. 4(a)). For $B_y \leq 6$ T, $R_{xx}^{2\omega}$ monotonically decreases with temperature and vanishes at around 30 K, which is close to the ferromagnetic transition temperature of this sample. On the other hand, the temperature dependence of $R_{xx}^{2\omega}$ for $B_y \geq 8$ T shows a characteristic hump as indicated by a down arrow. In addition, the peak temperature of the hump increases with $B_y$ (Fig. 4(b)). Since the excitation gap of the magnon increases with magnetic field, $R_{xx}^{2\omega}$ mediated by magnon decreases at high magnetic field and low temperature. This behavior also supports the magnon-mediated scattering mechanism in (Ge,Mn)Te. The similar temperature and magnetic field dependences were observed in the magnetic topological insulator bilayer [15].

We next investigate the $E_F$ dependence of the nonreciprocal response. We can successfully control the hole density in the sample from $p = 1.8 \times 10^{19}$ to $1.1 \times 10^{21}$ cm$^{-3}$ by varying the amount of supplied Te during the thin film growth (see Section S3 in SM). There exists three $E_F$ regions that have different types of the Fermi surface, which we label as (i), (ii), and (iii) as shown in Fig. 5(a). In region (i), there is a single Fermi surface. On the other hand, there are two Fermi surfaces in regions (ii) and (iii). The direction of the spin for two surfaces in regions (ii) and (iii) are the same and opposite, respectively. According to the model calculation (see section S4 in SM), $E_F$ in our samples ranges from the region (i) to around the boundary between (ii) and (iii). Figure 5(b) shows the $p$-dependence of the coefficient of nonreciprocal electrical transport $\gamma$'; the magnetic field dependences of $R_{xx}^{2\omega}$ for all the samples are shown in Fig. S5 in SM. One can see the significant enhancement of $\gamma$' in samples with small $p$. This dependence is consistent with the scenario of magnon-mediated scattering. Because of a small and single Fermi surface in the small $p$ regime (region (i)), magnons with small wave number and low energy dominantly contribute to electron scattering, which are easily populated even at low temperatures. On the other hand, the suppression of $\gamma$' in the large $p$ regime can be understood by the presence of two spin-polarized Fermi surfaces at $E_F$ (regions (ii) and (iii)). The complicated spin structure on the Fermi surface increases the scattering rate without spin flip, which reduces the ratio of the magnon-mediated scattering to the total scattering process. Thus, the exchange-field induced deformation of the Rashba band plays an important role to realize the single spin-momentum-locked Fermi surface where the magnon scattering mechanism works most effectively.

It is noteworthy that there may be another possible mechanism for the nonreciprocal electrical resistance other than the magnon-assisted spin scattering process. As demonstrated in a bulk Rashba semiconductor BiTeBr, the elastic scatterings on the deformed Fermi surface by in-plane magnetic field produces the nonreciprocal resistance [27]. However, the elastic mechanism alone cannot account for the magnetic field dependence of the observed nonreciprocal resistance, because the nonreciprocal resistance caused by the elastic scattering is linear to magnetic field. It may play some role, especially in high magnetic field region, where the magnon scattering is suppressed.

In summary, we have investigated the nonreciprocal electrical transport in multiferroic semiconductor (Ge,Mn)Te thin films. The magnon-associated spin scattering of electron on the spin-polarized Fermi surface causes the characteristic temperature and magnetic field dependences of nonreciprocal resistance. In addition, as a consequence of the deformation of the Rashba band induced by exchange field from the ferromagnetically ordered Mn moments, the single spin-momentum-locked Fermi surface appears in low hole density region, which significantly enhances the nonreciprocal electrical transport. Our observation of the nonreciprocal electrical transport in the magnetic Rashba system (Ge,Mn)Te paves a way to explore novel exotic phenomena in conducting multiferroic materials with both broken symmetries of time-reversal and space-inversion.

**FIGURE LEGENDS**

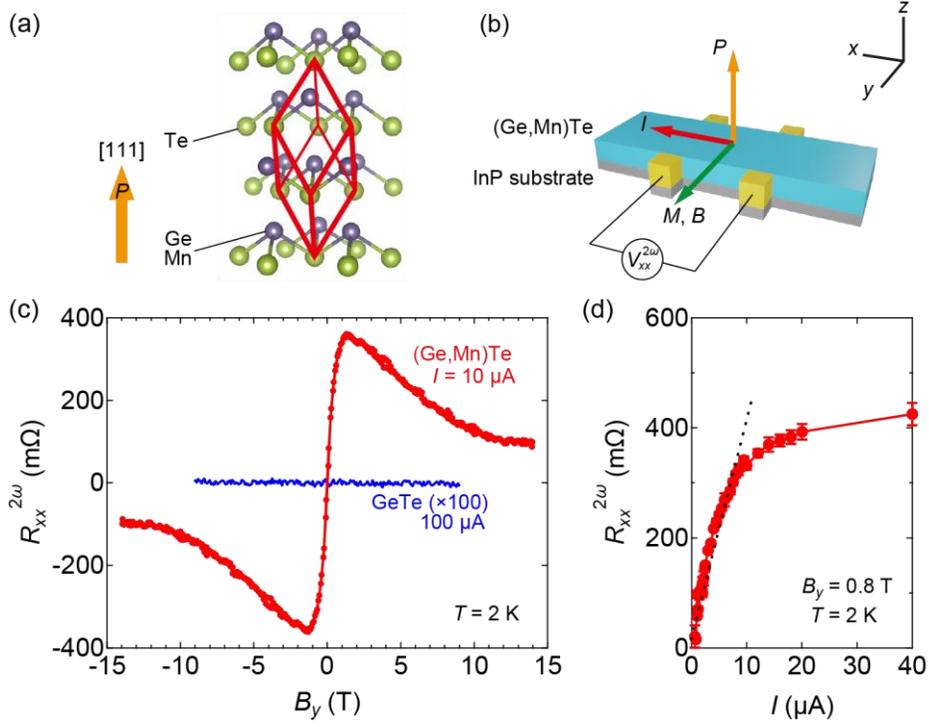

Figure 1 (a) Crystal structure of (Ge,Mn)Te with polar distortion along [111] direction. Red lines represent the unit cell. (b) A schematic of measurement configuration for nonreciprocal electrical transport in (Ge,Mn)Te thin films. (c) In plane magnetic field $B_y$ dependence of nonreciprocal resistance $R_{xx}^{2\omega}$ for (Ge,Mn)Te and GeTe thin films measured at temperature $T = 2$ K. The excitation current is 10 µA for (Ge,Mn)Te and 100 µA for GeTe. (d) Current dependence of $R_{xx}^{2\omega}$ of (Ge,Mn)Te at $T = 2$ K and $B_y = 0.8$ T.

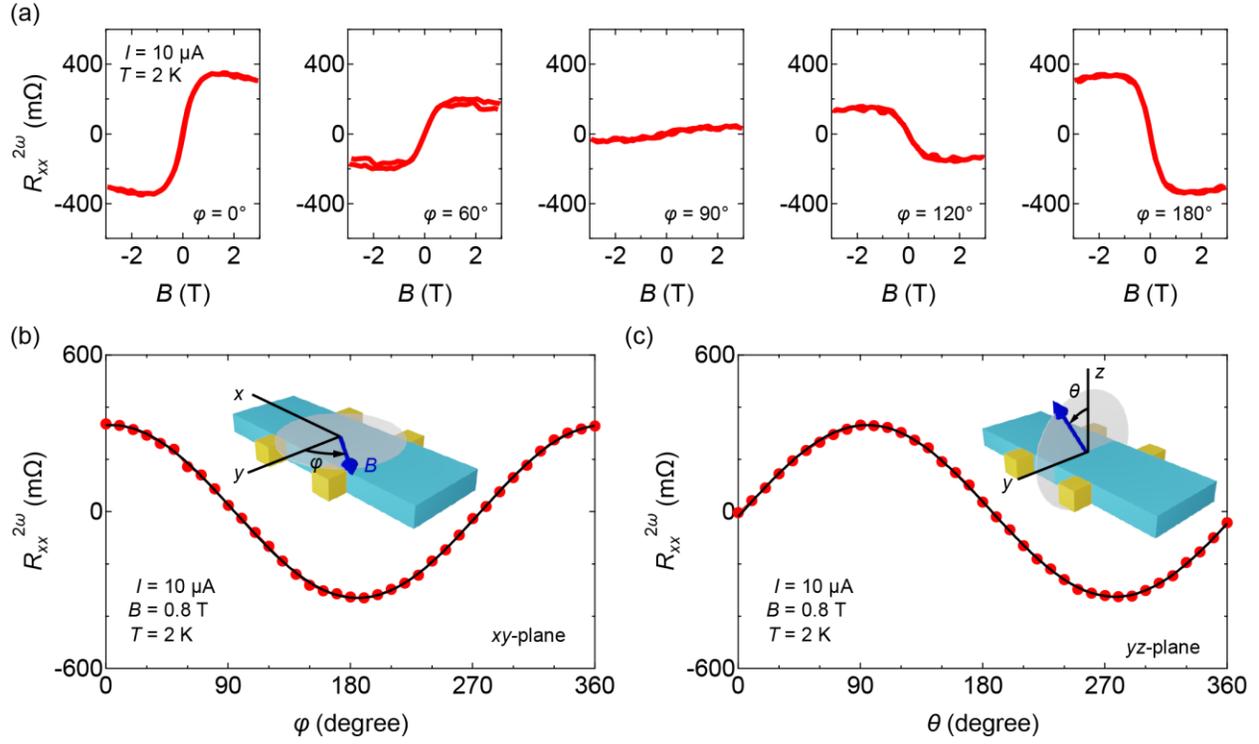

Figure 2 (a) Magnetic field dependence of nonreciprocal resistance $R_{xx}^{2\omega}$ for several azimuth angles, $\varphi = 0°, 60°, 90°, 120°$ and $180°$. (b), (c) In-plane (b) and out-of-plane (c) magnetic-field directional dependence of $R_{xx}^{2\omega}$ at $B = 0.8$ T and $T = 2$ K. Definition of in-plane and out-of-plane tilt angle $\varphi$ and $\theta$ are shown in the inset of (b) and (c), respectively.

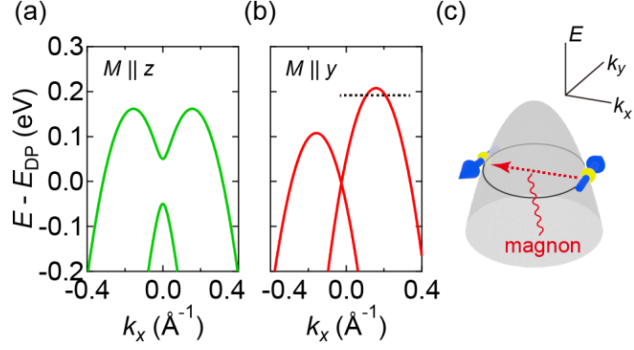

Figure 3 (a), (b) The band structures of (Ge,Mn)Te calculated by the model Hamiltonian (Eq. (2)) and parameters $m^* = 0.6\, m_0$, $\alpha = 2.0$ eVÅ, and $\Delta = 50$ meV for (a) out-of-plane ($M \parallel z$) and (b) in-plane ($M \parallel y$) magnetization. $E_{DP}$ represents the energy of the band degeneracy point without magnetization. A broken line in (b) represents the Fermi level for hole density $p = 1.8 \times 10^{19}$ cm$^{-3}$.

(c) A schematic of magnon-assisted spin scattering in a spin-momentum locked Fermi surface.

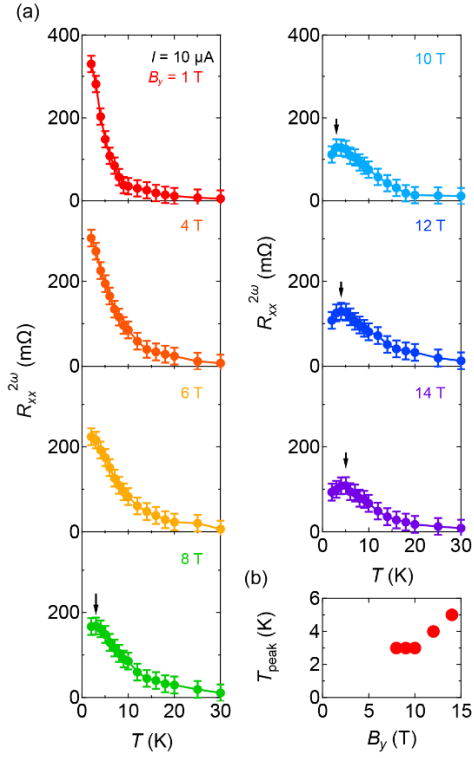

Figure 4 (a) Temperature dependence of nonreciprocal resistance $R_{xx}^{2\omega}$ under magnetic field $B_y$ = 1, 4, 6, 8, 10, 12 and 14 T. The black arrows in the panels of 8, 10, 12 and 14 T show the peak temperature $T_{peak}$. (b) Magnetic field $B_y$ dependence of $T_{peak}$.

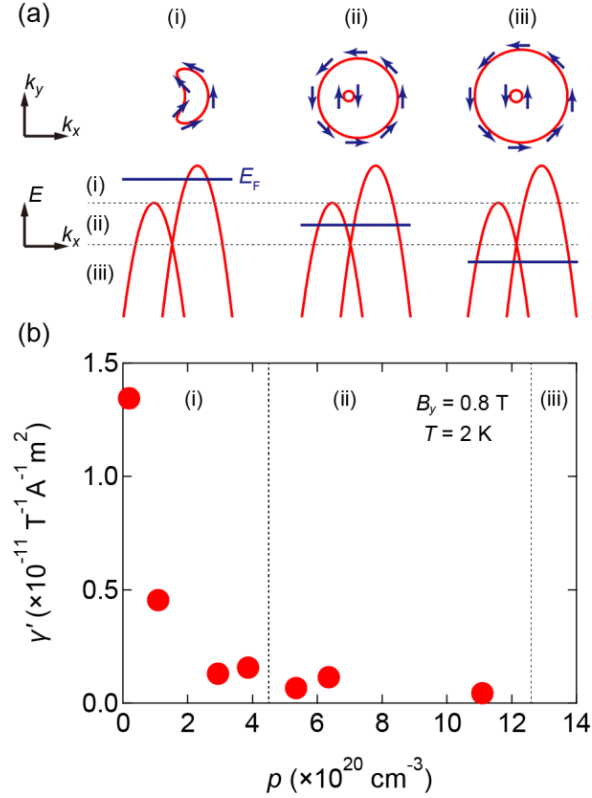

Figure 5 (a) Three types of the spin-momentum locked Fermi surfaces (top) that appear depending on the position of the Fermi level $E_F$ (bottom). (b) The coefficient of nonreciprocal electrical transport $\gamma'$ for seven (Ge,Mn)Te samples with different hole density $p$.

## ACKNOWLEDGMENTS

This work was partly supported by the Japan Society for the Promotion of Science through JSPS/MEXT Grant-in-Aid for Scientific Research (No.18H01155), and JST CREST (Nos. JPMJCR16F1, JPMJCR1874).

Ryutaro Yoshimi[1*], Minoru Kawamura[1], Kenji Yasuda[2], Atsushi Tsukazaki[3],

Kei S. Takahashi[1], Masashi Kawasaki[1, 4], and Yoshinori Tokura[1, 4, 5]

[1] *RIKEN Center for Emergent Matter Science (CEMS), Wako 351-0198, Japan.*

[2] *Department of Physics, Massachusetts Institute of Technology, Cambridge, Massachusetts 02139, USA*

[3] *Institute for Materials Research, Tohoku University, Sendai 980-8577, Japan*

[4] *Department of Applied Physics and Quantum-Phase Electronics Center (QPEC), University of Tokyo, Tokyo 113-8656, Japan*

[5] *Tokyo College, University of Tokyo, Tokyo 113-8656, Japan*

\* Corresponding author: ryutaro.yoshimi@riken.jp

## S1. Evaluation of nonreciprocal coefficient in pristine GeTe

Figure S1(a) shows the magnetic field dependence of nonreciprocal resistance $R_{xx}^{2\omega}$ for the GeTe thin film with excitation current of $I$ = 100, 500, 1000, 1500, 2000, and 2500 μA. The data for 100 μA is identical to that shown in Fig. 1(c) in the main text. The linear magnetic field dependence of $R_{xx}^{2\omega}$ is discernible with high excitation current. Figure S1(b) shows the excitation current dependence of $R_{xx}^{2\omega}$ at magnetic field $B$ = 9 T. $R_{xx}^{2\omega}$ liniearly increases with excitation current. We evaluate nonreciprocal coefficient $\gamma$' from these magnetic field and current dependences to be $\gamma$' = 1.4 × $10^{-16}$ $m^2T^{-1}A^{-1}$. This value is as large as that in the previous literature (~ 2-5 × $10^{-16}$ $m^2T^{-1}A^{-1}$) [28], yet five orders of magnitude smaller than that in (Ge,Mn)Te (1.3 × $10^{-11}$ $m^2T^{-1}A^{-1}$).

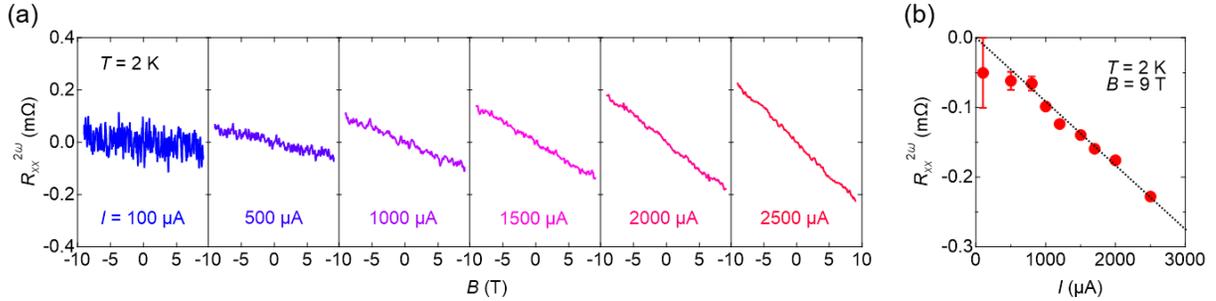

Figure S1 (a) Magnetic field dependence of nonreciprocal resistance $R_{xx}^{2\omega}$ for the pristine GeTe thin film with excitation current $I$ = 100, 500, 1000, 1500, 2000, and 2500 μA. (b) $I$-dependence of $R_{xx}^{2\omega}$ at $B$ = 9 T.

## S2. Magnetic field dependence of nonreciprocal resistance $R_{xx}^{2\omega}$ with different excitation current

Figure S2 shows the magnetic field dependence of nonreciprocal resistance $R_{xx}^{2\omega}$ for the (Ge,Mn)Te sample with hole density $p = 1.8 \times 10^{19}$ cm$^{-3}$ with excitation current $I = 1, 2, 5, 10, 20$ and 40 µA. The magnitude of $R_{xx}^{2\omega}$ increases with excitation current (the $I$-dependence of $R_{xx}^{2\omega}$ at $B = 0.8$ T is shown in Fig. 1(d) in the main text). The magnetic field dependence changes at higher excitation current. In particular, the decrease in $R_{xx}^{2\omega}$ towards high magnetic field becomes unclear at $I = 20$ and 40 µA.

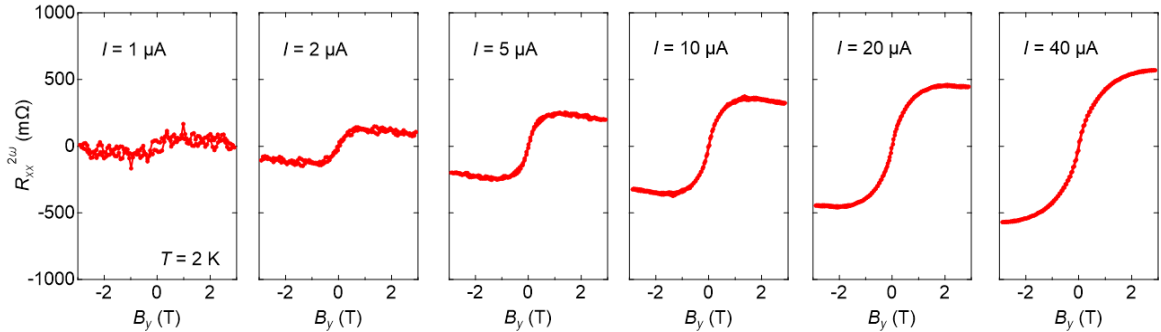

Fig. S2 Magnetic field dependence of nonreciprocal resistance $R_{xx}^{2\omega}$ for the (Ge,Mn)Te sample with hole density $p = 1.8 \times 10^{19}$ cm$^{-3}$ with excitation current $I = 1, 2, 5, 10, 20$ and 40 µA.

## S3. Transport properties of (Ge,Mn)Te with different growth conditions

We changed the hole density of (Ge,Mn)Te thin films by varying the equivalent beam pressure for Te ($P_{Te}$) during the MBE thin film growth. We fixed the equivalent beam pressure for Ge $P_{Ge}$ = 4.5 × 10$^{-6}$ Pa and changed that for Te from $P_{Te}$ = 2.0 × 10$^{-5}$ to 8.0 × 10$^{-5}$ Pa. Figure S3 shows the transport properties for samples with different $P_{Te}/P_{Ge}$. The temperature dependence of $\rho_{xx}$ becomes more metallic with increasing $P_{Te}/P_{Ge}$ (Fig. S3(a)). Accordingly, hole density $p$, evaluated by the ordinary Hall effect for $B > 1$ T, monotonically increases from $p$ = 1.8 × 10$^{19}$ cm$^{-3}$ to 1.1 × 10$^{21}$ cm$^{-3}$ (Fig. S3(b)). All the samples show ferromagnetic hysteresis at $T$ = 2 K.

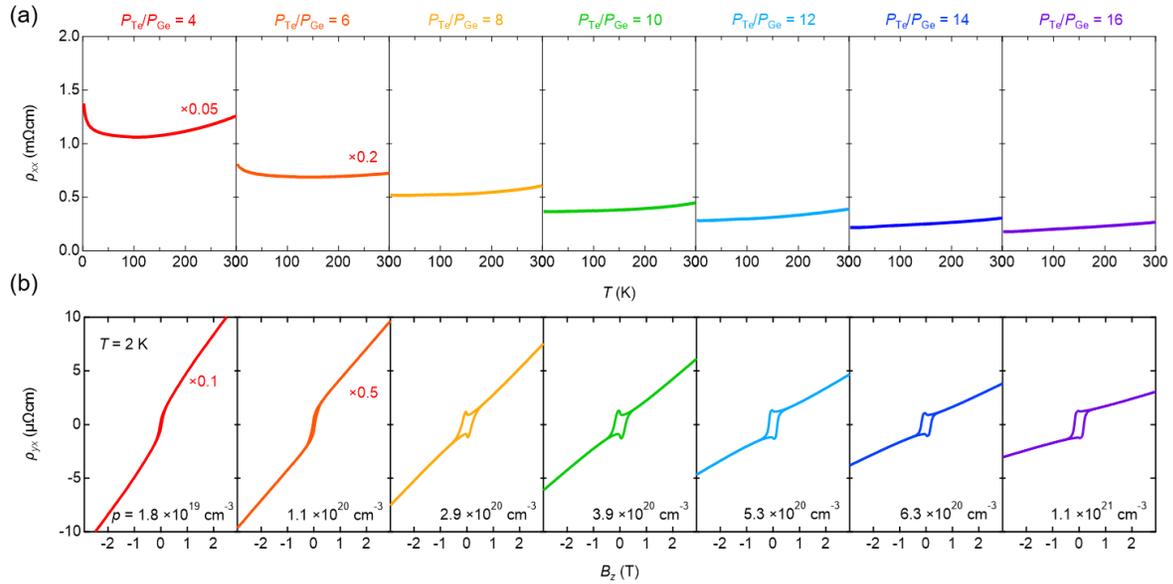

Fig. S3 (a) Temperature dependence of longitudinal resistivity $\rho_{xx}$ for (Ge,Mn)Te thin films grown with different supplied ratios of the equivalent beam pressures for Ge ($P_{Ge}$) and Te ($P_{Te}$). (b) Magnetic field dependence of Hall resistivity $\rho_{yx}$ at $T$ = 2 K.

## S4. The calculation of hole density as a function of $E_F$ in (Ge,Mn)Te

We calculate the band structure of (Ge,Mn)Te using the Hamiltonian (Eq. (2) in the main text) and $m^* = 0.6\ m_0$, $\alpha = 2.0$ eVÅ, and $\Delta = 50$ meV. The band structures are shown in Figs. S4(a) and (b) for out-of-plane ($M \parallel z$) and in-plane ($M \parallel y$) magnetization, respectively. We also calculate the Fermi level $E_F$ dependence of hole density $p$ (Fig. S4(c)). In the present experiment, the $p$ ranges from $1.8 \times 10^{19}$ to $1.1 \times 10^{21}$ cm$^{-3}$, which corresponds to the regions (i), (ii), and a small region in (iii) if it enters.

It is noteworthy that the actual constant energy surface in (Ge,Mn)Te has a hexagonal distortion that is represented by a $k^3$-term [22]. We do not adopt the term in the present model Hamiltonian, because we only think about the band structure near the band edge where $k$ is small.

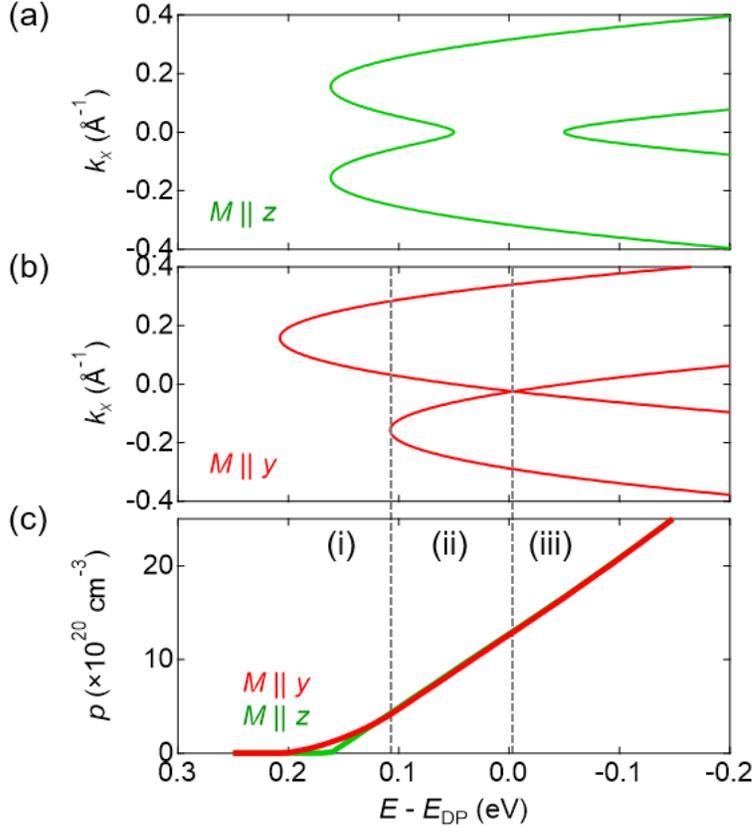

Fig. S4 (a), (b) The band structures of (Ge,Mn)Te calculated by the model Hamiltonian (Eq. (2) in the main text) and parameters $m^* = 0.6\ m_0$, $\alpha = 2.0$ eVÅ, and $\Delta = 50$ meV for (a) out-of-plane ($M \parallel z$) and (b) in-plane ($M \parallel y$) magnetization. (c) The $E_F$ dependence of hole density $p$ for out-of-plane and in-plane magnetization. $E_{DP}$ represents the energy of the band degeneracy point without magnetization.

## S5. Nonreciprocal electrical transport for samples with different hole densities

Figure S5 shows the magnetic field dependence of nonreciprocal resistance $R_{xx}^{2\omega}$ for (Ge,Mn)Te samples with different hole densities. In order to measure $R_{xx}^{2\omega}$ in the low excitation regime where $R_{xx}^{2\omega}$ is linear to the current, we set different excitation currents $I$ for samples. All the samples show a similar magnetic field dependence of $R_{xx}^{2\omega}$; a steep increase below $B \sim 0.5\text{-}1$ T and decrease in higher $B$ region. The similar magnetic field dependence regardless of hole density indicates that the nonreciprocal electrical transport originates from the common mechanism.

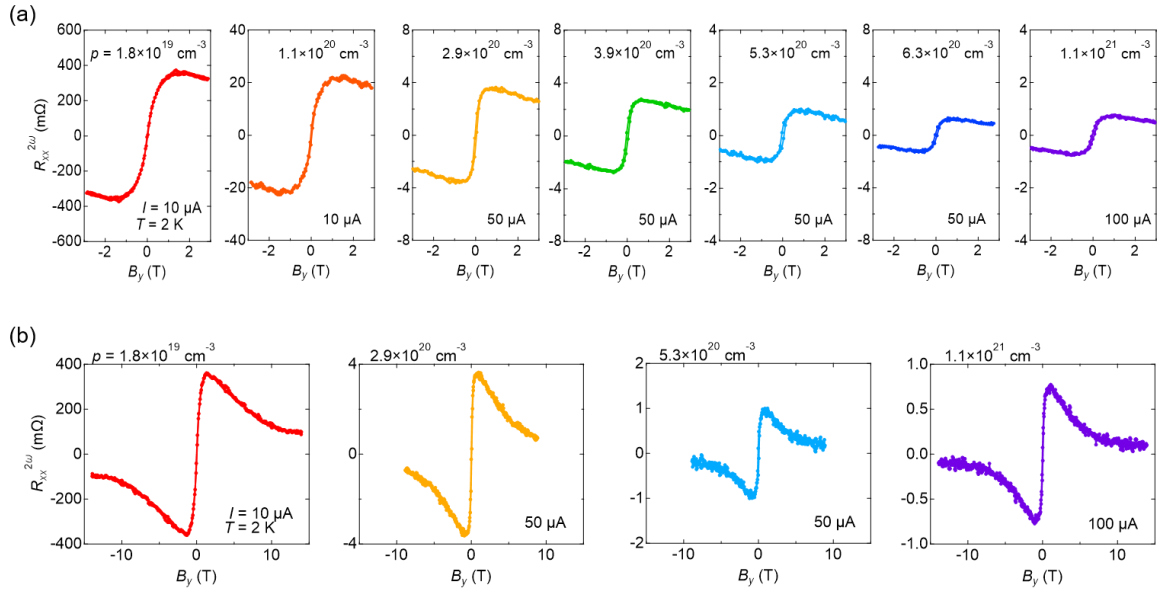

Fig. S5 (a) The magnetic field $B$ dependence of nonreciprocal resistance $R_{xx}^{2\omega}$ for (Ge,Mn)Te for samples with different hole density $p$. (b) The same $B$ dependence of $R_{xx}^{2\omega}$ for a wider magnetic field range for samples with $p = 1.8 \times 10^{19}$, $2.9 \times 10^{20}$, $5.3 \times 10^{20}$ and $1.1 \times 10^{21}$ cm$^{-3}$.